\documentclass[nofootinbib,aps,prab,reprint,groupedaddress,showpacs]{revtex4-2}
\usepackage[english]{babel}

\usepackage[utf8x]{inputenc}
\usepackage{amssymb}
\usepackage{amsthm}
\usepackage{amsmath}

\usepackage{hyperref}
\usepackage{bm}

\hypersetup{
    colorlinks=true, 
    linkcolor=cyan,
    citecolor=magenta, 
    filecolor=magenta, 
    urlcolor=cyan,
    runcolor=cyan
}

\usepackage{picture}
\usepackage{color,soul}
\usepackage{nth}
\usepackage{gensymb}
\usepackage{cleveref}
\usepackage{textcomp}
\usepackage{makecell}
\usepackage{lineno} 
\usepackage{textcomp}
\usepackage{siunitx}

\usepackage{comment}

\usepackage{bm}
\usepackage{physics}

\usepackage{upgreek}
\usepackage{mathtools}
\usepackage{subcaption}

\usepackage[english]{babel}

\makeatletter
\renewcommand{\fnum@figure}{FIG. \thefigure}
\makeatother

\begin{document}

\title{Plasma Discharge Undulator: a novel concept for plasma-based radiation sources}

\author{A. Frazzitta$^{1,2,3}$}

\affiliation{
$^{1}$Department of Physics, University of Rome “La Sapienza”, p.le A. Moro, 2-00185 Rome, Italy\\
$^{2}$Department of Physics, INFN—Milan, Via Celoria, 16-20133 Milan, Italy}


\date{\today}

\begin{abstract}
Plasma discharge devices have recently emerged as compact and versatile tools for particle beam manipulation. Building upon the Active Plasma Lens (APL) and its curved extension, the Active Plasma Bending (ABP), this work introduces the concept of the Plasma Discharge Undulator (PDU). In a PDU, a high-current discharge within a capillary generates an azimuthal magnetic field providing strong linear focusing (\(O(\mathrm{kT/m})\)), while a controlled and periodical spatial modulation of the discharge axis acts as a geometric driving term. The resulting beam dynamics can be modeled as a forced harmonic oscillator, yielding a well-defined oscillation at wavelength \(\lambda_{\mathrm{PDU}}\), distinct from the natural betatron wavelength \(\lambda_\beta\) related to APL focusing. Proper injection conditions result in the suppression of collective betatron oscillations, significantly reducing the intrinsic undulator strength spread typical of conventional plasma undulators, while allowing for matched beam transport thanks to APL strong focusing. Analytical models for particle trajectories and radiation emission are developed, and the one-dimensional requirements for free-electron laser (FEL) emission are evaluated, providing scaling relations and feasibility criteria for FEL operation in the proposed scheme. Theoretical estimates and multi-particle simulations indicate that the PDU can operate in the short-period regime (\(\lambda_{\mathrm{PDU}} = \mathrm{mm{-}cm}\)) with tunable undulator strength \(K_{\mathrm{PDU}}\), supporting narrow-band radiation emission. The PDU thus provides a pathway toward miniaturized, tunable, fully-plasma-based light sources with enhanced control over focusing and spectral properties. 
\end{abstract}

\maketitle

\section{Introduction}
Conventional plasma undulators (CPUs) \cite{Kostyukov2003,Esarey2002,betatron1,betatron2,Frazzitta2023,Galdenzi2024} have attracted considerable interest due to their intrinsically short undulation period, which, for a given beam energy, leads to higher radiation intensity and frequency compared to conventional undulators (CUs). CPUs are also currently under investigation as potential compact free-electron lasing (FEL) systems \cite{Whittum1990,Litos2018,Davoine2018,Chen1990}. A common feature of these devices is that the undulator period \(\lambda_U\) is replaced by the betatron wavelength \(\lambda_\beta \propto \sqrt{\gamma / n_p}\), with \(\gamma\) as the beam's Lorentz factor and \(n_p\) the plasma density. A feature related to this replacement is that, within a particle beam, each element exhibits a different oscillation amplitude \(x_0\), and consequently a different undulator strength parameter \(K = x_0 \gamma k_\beta\) \cite{jackson}, with \(k_\beta\) as the betatron wavenumber. For a Gaussian beam with any transverse size \(\sigma_r\) injected on-axis into the electrostatic linear focusing field typical of the blowout regime \cite{Pukhov2002,Esarey2009}, it can be shown that the corresponding undulator strength spread is fundamentally limited by \(\sigma_K / K \geq 0.523\) for a flat matched Gaussian beam and \(\sigma_K / K \geq 0.363\) for a symmetric matched round Gaussian beam. One possible way to mitigate this intrinsic spread is to inject the beam centroid off-axis, thereby inducing a collective oscillation. However, unless the offset exceeds approximately \(3\sigma_r\), some particles will remain in counter-phase; moreover, this approach may also excite hosing instabilities \cite{Huang2007} or damp out toward the axis, leading to a growth of transverse emittance, especially in the case of long propagation distances and/or multiple plasma stages. Furthermore, an additional independent source of \(K\)-spread arises from the beam energy spread \(\sigma_\gamma / \gamma\), which affects both the betatron wavelength and the relativistic Doppler shift, thereby broadening the emitted spectrum.

\begin{figure}[t]
    \centering
    \includegraphics[width=0.99\linewidth]{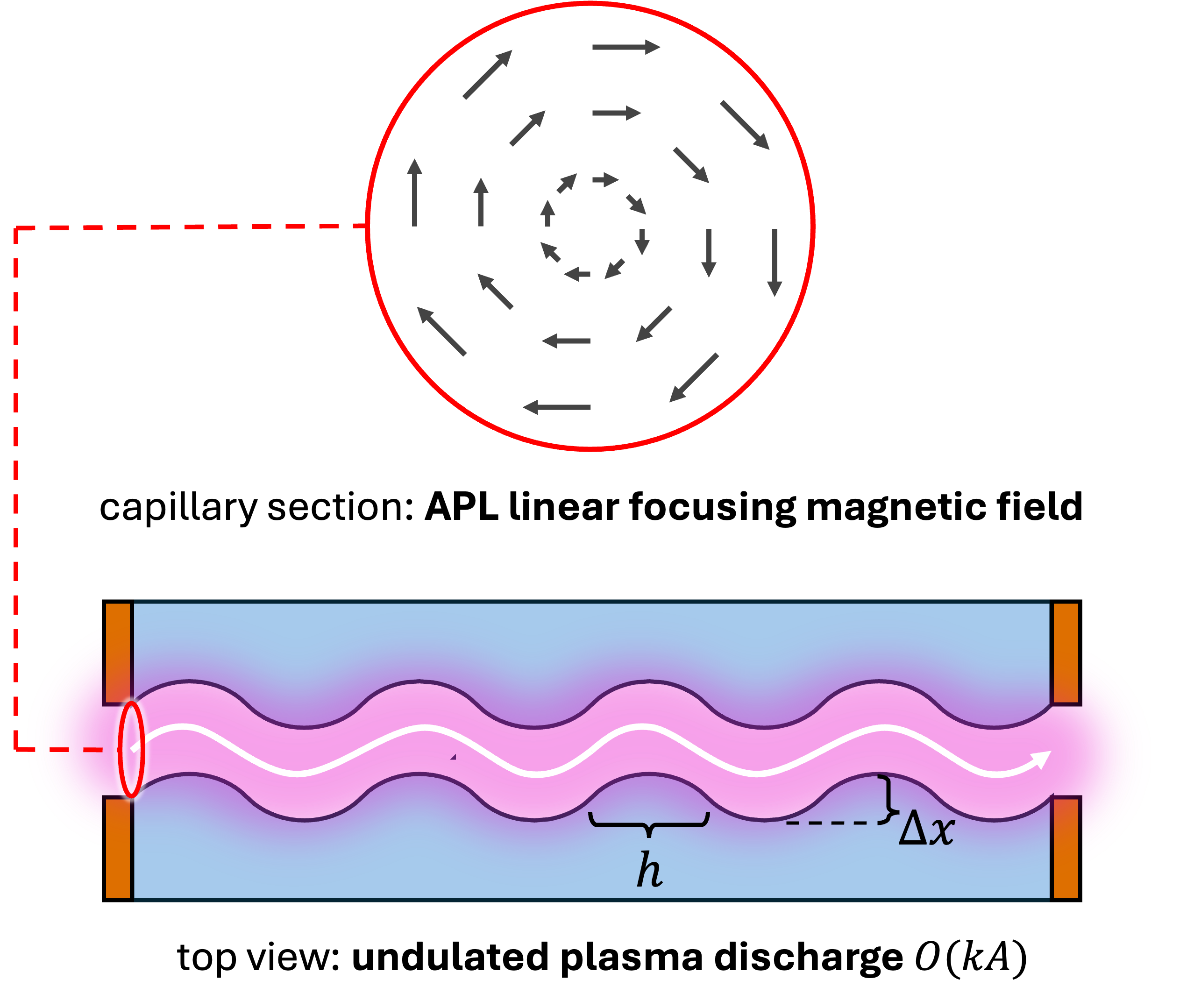}
    \caption{Conceptual sketch of the Plasma Discharge Undulator (PDU). Top: azimuthal magnetic field distribution induced by the discharge current in a capillary. Bottom: example of possible geometrical implementation of the periodic discharge modulation.}
    \label{PDUcartoon}
\end{figure}

A key feature of CPUs leading to these intrinsic limitations is the absence of an external driving term in the oscillatory motion. In this work, a method is proposed to introduce such a forcing term, thereby stabilizing the emission process, while also freeing the scheme from the need for a high-energy-density driver such as an intense laser pulse or a particle beam: the \textit{Plasma Discharge Undulator} (PDU). In this concept for a plasma-based undulator, the oscillating and focusing field is magnetic in nature, rather than electrostatic as in CPUs: building upon the principles of the \textit{Active Plasma Lens} (APL) \cite{vanTilborg2015,Pompili2018,pompili2018focusing} and \textit{Active Plasma Bending} (ABP) \cite{Pompili2024,Frazzitta2024}, the PDU relies on a high-current (\(O(1-10)\) kA) discharge within a capillary previously filled with neutral gas and subsequently ionized by applying a high voltage (\(O(10)\) kV). The associated magnetic field follows a Biot–Savart configuration, with azimuthal field lines around the capillary axis and a magnitude that increases linearly with radial distance (see Fig.~\ref{PDUcartoon}, top), providing a linear focusing force. The proposed PDU scheme consists in establishing a spatially sinusoidal (or possibly helical) modulation of the discharge current: the formation of such a modulated current profile would periodically displace the beam off-axis with respect to the local zero of the discharge-generated focusing magnetic field, thereby imposing an alternating dipolar kick that results in the overall undulating motion of the beam. Moreover, the intense transverse focusing (\(O(\mathrm{kT/m})\)) enables effective beam transport under matched APL-like conditions, providing an extra degree of control over the beam dynamics evolution—typically subject to drift or weak focusing in CUs, where matching is usually not achievable. One more advantage of the PDU configuration, now compared to CPUs, is that the undulator wavelength becomes independent of the beam energy and that the \(K\)-spread is intrinsically lower, both feature reducing the broadening of the emitted spectrum. 

Achieving the proposed current modulation is currently under investigation and could be realized through capillaries with sinusoidal or stepwise-offset geometries (see Fig.~\ref{PDUcartoon}, bottom). A favorable aspect of this approach is the extensive body of work developed in recent years on plasma discharge capillaries \cite{Biagioni2021,Crincoli2025,Arjmand2025}, accompanied by a growing interest in nonlinear channel configurations \cite{Pompili2024,Frazzitta2024}---of which the PDU represents a specific realization---and by the parallel ongoing development of numerical tools capable of reliably simulating these systems \cite{Mewes2025}, including thermal transport and boundary effects. There are also sound physical reasons to expect that the discharge current should follow, at least partially, a spatial profile imposed by nonlinear capillary walls. The first argument is of thermal nature: regardless of the current direction, the plasma temperature in a discharge capillary is minimized at the inner walls, which act as an effective heat sink \cite{Bobrova2001,Curcio2019}. This gives rise to a transverse temperature profile that encodes information about the wall geometry into the plasma volume. Since the electrical resistivity decreases with increasing temperature, the region of maximum temperature---and therefore minimum resistance---is expected to be influenced by the wall shape, favoring a current distribution correlated with the capillary geometry. A second argument concerns the nature of the current carriers in the plasma, namely electrons and ions, and their large mass disparity (exceeding a factor \(m_i/m_e\approx1836\)). Typical discharge capillary parameters, with current densities $J = O(10^9\text{--}10^{10})\,\mathrm{A/m^2}$ and plasma densities $n_p = O(10^{17})\,\mathrm{cm^{-3}}$, correspond to electron drift velocities $v_d = O(10^5)\,\mathrm{m/s}$. Over a characteristic discharge duration $\tau = O(1)\,\mu\mathrm{s}$, electrons can therefore travel distances of $O(1\text{--}10)\,\mathrm{cm}$, comparable with typical capillary lengths. Conversely, ions are several thousand times slower and can be regarded as effectively stationary within the capillary, remaining distributed according to the initial ionization profile, which for small capillaries typically fills the entire volume bounded by the walls. Any local charge separations relax on timescales of order $1/\omega_p \ll \tau$, effectively enforcing quasi-neutral plasma flow influenced by capillary geometry. This latter reasoning does not holds for possible negative charge accumulation at the capillary walls \cite{bohm1949}, which, if present, would further assist the spatial modulation of the current direction.

\section{Trajectories}
\begin{figure}[t]
\begin{subfigure}{0.8\linewidth}
\centering
\includegraphics[width=\linewidth]{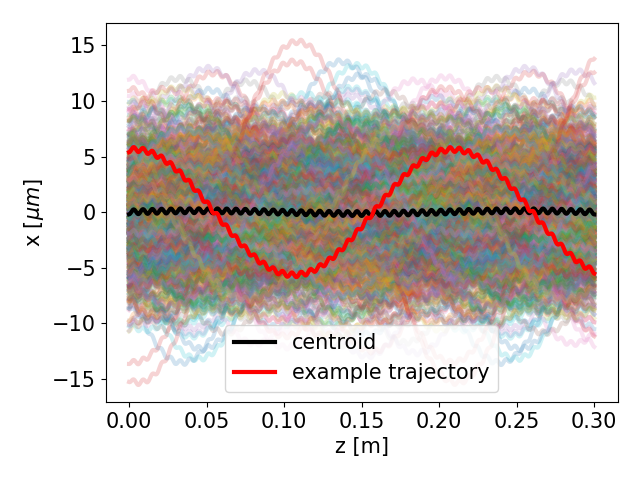}
\caption{}
\label{trajmatch}
\end{subfigure}
\begin{subfigure}{0.78\linewidth}
\centering
\includegraphics[width=\linewidth]{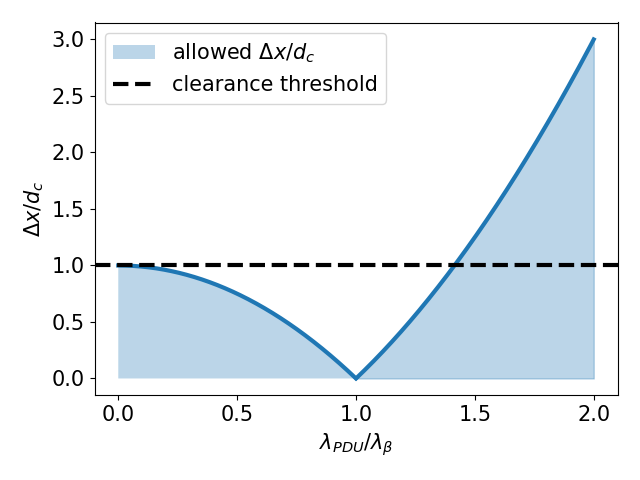}
\caption{}
\label{clearance}
\end{subfigure}
\caption{(a) Matched beam propagation in PDU field for proper centroid injection offset \(x_{inj}=x_0\), which provide collective oscillation at \(\lambda=\lambda_\mathrm{PDU}\) only (solid black). Single particle trajectories will be anyway subject to mixed betatron/undulator motion (solid red as an example) (b) Geometric constraints on capillary offset \(\Delta x\) normalized over capillary diameter \(d_c\) as a function of \(\lambda_{\mathrm{PDU}}/\lambda_\beta\), showing that for \(\lambda_{\mathrm{PDU}}<\lambda_{\mathrm{\beta}}\) the system necessarily needs clearance (i.e. a line of sight through the device).}
\label{dynamics}
\end{figure}

The single-particle trajectory model adopted for the Plasma Discharge Undulator (PDU) is a simple forced harmonic oscillator:
\begin{equation}
    \frac{d^2x}{dz^2} = -k_\beta^2 \big(x - x_{eq}(z)\big),
\end{equation}
where \(k_\beta\) is the betatron wavenumber determined by the magnetic focusing strength \cite{vanTilborg2015,Frazzitta2024},
\begin{equation}
    k_\beta = \sqrt{\frac{e \mu_0 J}{2 m_e c \gamma}},
\end{equation}
\(e\) as the elementary charge, \(\mu_0\) as the vacuum magnetic permeability, \(J\) as the current density, \(c\) as the speed of light and where \(x_{eq}(z)\) describes the spatial dependence of the magnetic field kernel imposed by the sinusoidal modulation of the discharge current:
\begin{equation}
    x_{eq}(z) = \frac{\Delta x}{2} \cos(k_{\mathrm{PDU}} z),
\end{equation}
where \(\Delta x/2\) is the oscillation amplitude, \(k_{\mathrm{PDU}} = \pi / h\) is the undulation wavenumber and \(h = \lambda_{\mathrm{PDU}}/2\) is the half-period of the modulation (see Figure~\ref{PDUcartoon}, bottom). The general solution for the transverse trajectory \(x(z)\) is
\begin{equation}
\begin{aligned}
    x(z) = & \,\,x_0 \cos(k_\beta z) +\\
    + \frac{\Delta x}{2} \frac{k_\beta^2}{k_\beta^2 - k_{\mathrm{PDU}}^2}&
    \big[\cos(k_{\mathrm{PDU}} z) - \cos(k_\beta z)\big],
\end{aligned}
\end{equation}
which represents a superposition of oscillations at the betatron wavelength \(\lambda_\beta\) and at the undulation wavelength \(\lambda_{\mathrm{PDU}}\). By choosing the initial amplitude \(x_0\) such that the \(\lambda_\beta\) oscillation term is canceled, one obtains
\begin{equation}\label{x0}
    x_0 = \frac{\Delta x}{2} \frac{k_\beta^2}{k_\beta^2 - k_{\mathrm{PDU}}^2}\coloneqq x_{off} 
\end{equation}
Injecting the beam centroid at \(x = x_{off}\) therefore eliminates any collective centroid oscillation at \(\lambda = \lambda_\beta\), leaving only the motion at \(\lambda = \lambda_{\mathrm{PDU}}\) (see Fig.~\ref{trajmatch}, solid black). Even under these conditions, individual particles still possess a finite transverse distribution, resulting in random injection amplitudes and in general giving an oscillatory motion which superposes \(\lambda_\beta\) and \(\lambda_\mathrm{PDU}\) (see an highlighted example in Fig.~\ref{trajmatch}, solid red). 

It is worth noting that, for the special case \(\lambda_{\mathrm{PDU}} = \lambda_\beta\), the initial offset \(x_{off}\) diverges (theoretical resonance), while it becomes negative (positive) for \(\lambda_{\mathrm{PDU}} < (>) \lambda_\beta\). This behavior defines important design guidelines for the device: in the favorable regime in terms of photon energy (i.e. \(\lambda_{\mathrm{PDU}} < \lambda_\beta\)), particles undulate in counter-phase with respect to the geometric forcing, thus \emph{requiring} a finite clearance through the device. Imposing the condition that the beam centroid must not touch the capillary wall yields:
\begin{equation}
    \Delta x <
  \frac{2r_c}{\dfrac{k_\beta^2}{k_{\mathrm{PDU}}^2 - k_\beta^2} + 1} 
    \qquad \text{for } \lambda_{\mathrm{PDU}} < \lambda_\beta,
\end{equation}
and
\begin{equation}
    \Delta x <
  \frac{2r_c}{\dfrac{k_\beta^2}{k_\beta^2 - k_{\mathrm{PDU}}^2} - 1} 
    \qquad \text{for } \lambda_{\mathrm{PDU}} > \lambda_\beta.
\end{equation}
with \(r_c\) as the circular capillary section's radius. In the former case, the total offset between adjacent sections always satisfies \(\Delta x < 2r_c\) (forced clearance), while in the latter, the forced clearance condition \(\Delta x < 2r_c\) applies for \(\lambda_\beta < \lambda_{\mathrm{PDU}} < \sqrt{2}\lambda_\beta\), and a no-clearance configuration (\(\Delta x > 2r_c\)) becomes possible for \(\lambda_{\mathrm{PDU}} > \sqrt{2}\lambda_\beta\) (see Fig.~\ref{clearance}).

Independently from beam centroid collective oscillations, the beam can be injected in a transversly matched condition following the standard relation for an APL \cite{vanTilborg2015,Frazzitta2024} (see Fig.~\ref{trajmatch}, where a matched beam transverse envelope evolution is presented):
\begin{equation}
    \sigma_M = 
    \left(
    \frac{2 m_e c \gamma \epsilon_{\mathrm{rms}}^2}
         {e \mu_0 J}
    \right)^{1/4},
\end{equation}
where \(\epsilon_{\mathrm{rms}}\) is the geometric root-mean-square emittance.

\section{Radiation}
The radiative properties of the PDU are directly determined by the features of particle trajectories. The primary parameter defining the emission features in an undulating system is the undulator strength parameter $K$. In the following subsections, simple limiting expressions for PDU undulator strength are first introduced, corresponding to the pure CU and CPU cases respectively. These limits are then unified through the derivation of a general expression for the effective undulator strength in the PDU configuration.

\subsection{PDU undulator strength}
In a PDU, the oscillation amplitude is given by \(x_{off}\) from Eq.~\eqref{x0}, and the wavenumber corresponds to \(k_{\mathrm{PDU}}\). Substituting the relevant quantities in the typical undulator strength expression \cite{jackson,Ciocci2000} yields
\begin{equation}\label{Kpdu}
    K_{\mathrm{PDU}} = x_{off} \gamma k_{\mathrm{PDU}} 
    = \frac{\gamma \Delta x}
           {\dfrac{2h}{\pi} - \dfrac{4\pi m_e c \gamma}{e \mu_0 J h}},
\end{equation}
which should be always taken in absolute value. Equation~\eqref{Kpdu} shows that this expression of the undulator strength grows linearly with the capillary amplitude offset \(\Delta x\). In the regime \(\lambda_{\mathrm{PDU}} < \lambda_\beta\), the second term in the denominator is dominant, leading to a linear dependence \(K_{\mathrm{PDU}} \propto h J \Delta x\),  and removing any dependance on \(\gamma\). Conversely, for \(\lambda_{\mathrm{PDU}} > \lambda_\beta\), one obtains \(K_{\mathrm{PDU}} \propto \gamma \Delta x / h\).
\begin{figure}[t]
    \centering
    \includegraphics[width=0.99\linewidth]{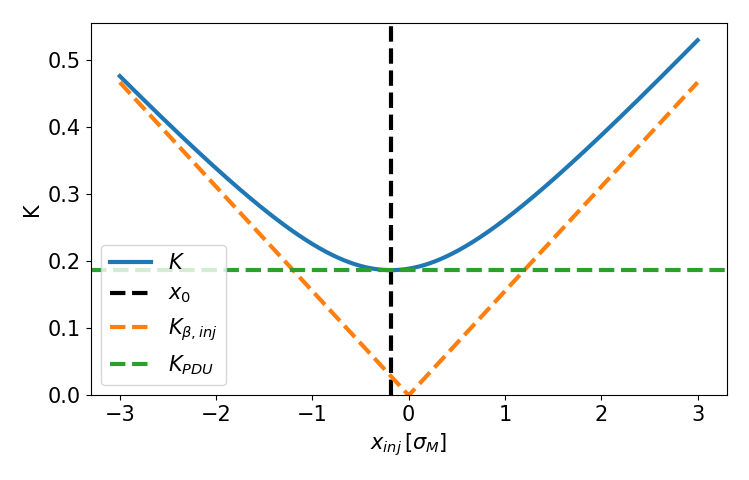}
    \caption{Comparison between the two asymptotic regimes of the single particle undulator strength \(K\) as a function of the transverse injection position in the plasma channel. For small amplitudes, \(K\) converges to the forced value \(K_{\mathrm{PDU}}\); for increasing amplitudes, it approaches the betatronic limit \(K_{\beta,inj} = r_{inj}\gamma k_\beta\), following a hyperbolic trend.}
    \label{Kcomp}
\end{figure}

\subsection{Betatron undulator strength}
An interesting feature of the PDU is its inherently two-color nature. Besides the forced \(\lambda_\mathrm{PDU}\) oscillation, the fundamental \(\lambda_\beta\) betatron oscillation remains present. Although this component does not reach the sub-millimeter wavelengths typical of CPUs \cite{Frazzitta2023,Galdenzi2024,Stocchi2025}, the strong magnetic focusing fields involved yield betatron wavelengths down to a few centimeters long. Consequently, the PDU exhibits two reference fundamental emission wavelengths and undulator strength parameter expression, \((\lambda_{1,\mathrm{PDU}},\, K_{\mathrm{PDU}})\) and \((\lambda_{1,\beta},\, K_\beta)\), where \(K_\beta\) is associated with the average oscillation amplitude within a Gaussian transverse distribution:
\begin{equation}
\begin{aligned}
        K_{\beta,ave} &= \sqrt{\frac{9\pi}{8}} \, \sigma_r k_\beta \gamma\\
        K_{\beta,mode} &= \sigma_r k_\beta \gamma\coloneqq K_{\beta}\\
\end{aligned}
\end{equation}
assuming again a matched beam condition.

\subsection{Effective undulator strength}
The two expressions for the undulator strength introduced above, $K_{\mathrm{PDU}}$ and $K_\beta$, can actually be shown to represent limiting cases of a more general formulation. In particular, it can be demonstrated~\cite{Schmser2009} that for a single trajectory the undulator strength parameter $K$ is \emph{unique} and fundamentally defined by the longitudinal component of the particle velocity:
\begin{equation}
    K = \sqrt{(1 - \overline{\beta_z})4\gamma^2 - 2}.
\end{equation}
Under the paraxial approximation, the complete expression of \(K\) for a single particle propagating in the PDU field becomes
\begin{equation}
    K \approx \gamma \sqrt{k_\beta^2 r_{inj}^2 + k_{\mathrm{PDU}}^2 x_0^2},
\end{equation}
where
\begin{equation}
    r_{inj} = \sqrt{[(x_{\mathrm{inj}} - x_0)^2 + y_{\mathrm{inj}}^2] + \frac{(x_{\mathrm{inj}}^{\prime 2} + y_{\mathrm{inj}}^{\prime 2})}{k_\beta^2}},
\end{equation}
and \(x_{\mathrm{inj}},\, y_{\mathrm{inj}},\, x_{\mathrm{inj}}',\, y_{\mathrm{inj}}'\) are the particle’s transverse injection coordinates and slopes. Interestingly, and somewhat intuitively, \(K \to K_{\mathrm{PDU}}\) for vanishing betatron oscillation amplitude, when only the \(\lambda_\mathrm{PDU}\) oscillation survives; conversely, for increasing betatron amplitude \(K \to r_{inj}\gamma k_\beta\coloneqq K_{\beta,inj}\) with a hyperbolic trend, restoring the basic single particle betatronic undulator strength. An example of this behavior is shown in Figure~\ref{Kcomp}, where the PDU single particle undulator strength is plotted as a function of transverse injection amplitude. Extending the analysis to a Gaussian particle beam, the first step is to normalize the transverse coordinates over the matched beam size \(\sigma_M\), defining a new variable \(\tilde{r}_{inj}\) which is distributed as a Rayleigh \(\chi_4\) function. The resulting \(K\) distribution can be analytically evaluated for \(\lambda_{\mathrm{PDU}} < \lambda_\beta\), yielding the following average and standard deviation estimators:
\begin{equation}\label{Kestim}
    \begin{aligned}
        \mu_K &= \frac{3}{2}K_{\mathrm{PDU}} + \frac{\gamma\mathcal{B}}{4k_\beta\sigma_M},\\
        \sigma_K &= \sqrt{|K_{\mathrm{PDU}}^2 + 4K_\beta^2 - \mu_K^2|},
    \end{aligned}
\end{equation}
with
\begin{equation}
    \begin{aligned}
        \alpha &= -\frac{K_{\mathrm{PDU}}}{\sqrt{2}K_\beta},\\
        \mathcal{Z} &= \frac{3K_\beta^2 - K_{\mathrm{PDU}}^2}{\gamma^2},\\
        \mathcal{B} &= \sqrt{2\pi}\,\mathcal{Z}\,\exp[\alpha^2]\,(1+\mathrm{erf}[\alpha]).
    \end{aligned}
\end{equation}
By evaluating the limits for \(\alpha \to 0\) (\(K_{\mathrm{PDU}} \ll K_\beta\)) and \(\alpha \to -\infty\) (\(K_{\mathrm{PDU}} \gg K_\beta\)), one can assess the behavior of the PDU compared with a CPU. 

For \(K_{\mathrm{PDU}} \ll K_\beta\):
\begin{equation}\label{KlimKb}
    \begin{aligned}
        \mu_K &\to \sqrt{\frac{9\pi}{8}}\,K_\beta \equiv K_{\beta,\mathrm{ave}}\\
        \sigma_K &\to \sqrt{4 - \frac{9\pi}{8}}\,K_\beta \equiv \sigma_{K_\beta}\\
        \frac{\sigma_K}{\mu_K} &\to \sqrt{\frac{32}{9\pi} - 1} \approx 0.36
    \end{aligned}
\end{equation}

In contrast, for \(K_{\mathrm{PDU}} \gg K_\beta\):
\begin{equation}\label{KlimKPDU}
    \begin{aligned}
        \mu_K &\to K_{\mathrm{PDU}}\\
        \sigma_K &\to 0\\
        \frac{\sigma_K}{\mu_K} &\to 0
    \end{aligned}
\end{equation}
Equation~\eqref{KlimKPDU} highlights that the regime \(K_{\mathrm{PDU}} \gg K_\beta\) is particularly advantageous in terms of radiation quality, as it leads to a vanishing \(K\)-spread for suitable beam parameters. Assuming all other beam and device parameters fixed, the condition of Eq.~\eqref{KlimKPDU} is satisfied for a normalized emittance
\begin{equation}\label{epsPDU}
    \epsilon_n \ll
    \left(\frac{I}{\tilde{I}_0}\right)^{3/2}
    \frac{\lambda_{\mathrm{PDU}}^2 \Delta x^2}{16\pi^2 r_c^3 \sqrt{\gamma}}\coloneqq\epsilon_\mathrm{PDU},
\end{equation}
where \(\tilde{I}_0 = 2\pi m_e c / \mu_0 e \approx 8.5~\mathrm{kA}\) is half the Alfvén current. For given beam and device parameters, less restrictive emittance conditions can therefore be achieved by increasing the discharge current and/or reducing the capillary radius, being \( \epsilon_\mathrm{PDU}\propto J^{3/2}\).

\subsection{PDU 1D FEL Conditions}
Relying on established results for free-electron laser (FEL) 1D lasing conditions~\cite{Schmser2009} together with the beam matching constraints within the PDU, it is possible to assess the feasibility of FEL processes in this configuration. Given the nature of the PDU device, the emittance constraint must include the effects of betatron motion, yielding~\cite{Schmser2009}:
\begin{equation}
    \epsilon_{\mathrm{rms}} < \frac{\beta_{\mathrm{ave}}}{2\sqrt{2}\gamma^2}\,\rho_{\mathrm{FEL}},
\end{equation}
where \(\beta_{\mathrm{ave}}\) is the average betatron function and \(\rho_{\mathrm{FEL}}\) is the Pierce parameter. Assuming a matched beam and recalling that \(\sigma_M = \sqrt{\epsilon_{\mathrm{rms}}\beta_{\mathrm{ave}}} = \sqrt{\epsilon_{\mathrm{rms}}/k_\beta}\), and that in the absence of energy spread \(\epsilon_n = \gamma \epsilon_{\mathrm{rms}}\), one obtains the following limit on the normalized emittance:
\begin{equation}\label{eps_FEL}
    \epsilon_n < \frac{\rho_{\mathrm{FEL}}}{2\sqrt{2}\gamma k_\beta}.
\end{equation}
Equation~\eqref{eps_FEL} can then be evaluated using the most appropriate expression for \(\rho_{\mathrm{FEL}}\) for the given case. For simplicity, the one-dimensional formulation reads~\cite{Schmser2009}:
\begin{equation}
    \rho_{\mathrm{FEL,1D}} = \frac{1}{4\pi\sqrt{3}}\frac{\lambda_{\mathrm{PDU}}}{L_{g,\mathrm{1D}}},
\end{equation}
where the 1D gain length is
\begin{equation}
    L_{g,\mathrm{1D}} = \frac{1}{\sqrt{3}}\left[\frac{4\gamma^3 m_e}{\mu_0 \mu_K^2 e^2 k_{\mathrm{PDU}} n_{\mathrm{beam}}}\right]^{1/3},
\end{equation}
with \(n_{\mathrm{beam}}\) denoting the beam number density and \(\mu_K\) the mean undulator strength from Eq.~\eqref{Kestim}. By estimating the matched beam density as \(n_{\mathrm{beam}} = Q / Ve\),  
where \(Q\) is the total bunch charge and \(V = 36\pi R \sigma_M^3\) is the 3-\(\sigma\) beam volume with \(R = \sigma_z / \sigma_M\) the beam aspect ratio,  
the emittance condition for the FEL instability growth becomes:
\begin{equation}\label{eps_FEL1}
    \epsilon_n \lessapprox \frac{2}{25}\frac{1}{\gamma k_\beta^{1/3}}\left(\frac{\mu_0\mu_K^2eQ}{m_ek_\mathrm{PDU}^2R}\right)^{2/9}\coloneqq\epsilon_{\mathrm{FEL}}.
\end{equation}
The two emittance constraints given by Eqs.~\eqref{epsPDU} and~\eqref{eps_FEL1} can be combined to yield the general condition:
\begin{equation}\label{epslims}
    \epsilon_n < \mathrm{min}\,(\eta\,\epsilon_{\mathrm{PDU}},\,\epsilon_{\mathrm{FEL}})
\end{equation}
with \(\eta\ll1\). The distinction between the regimes of validity of the two expressions can be expressed in terms of the bunch charge \(Q\) and the beam aspect ratio \(R\):
\begin{equation}
\begin{aligned}
    \left.\frac{Q}{R}\right|_{lim}=\frac{144\sqrt{2}\pi m_e}{\mu_0ek_\mathrm{PDU}^9}&\left(-\eta\gamma\Delta xk_\beta^{4/3}(k_\beta^2 - k_\mathrm{PDU}^2)\right)^{9/2}\times\\
    \times&\left(-\frac{\Delta x\gamma k_\beta^2}{k_\beta^2 - k_\mathrm{PDU}^2}\right)^{5/2}
\end{aligned}
\end{equation}
For \(Q/R < Q/R|_{\mathrm{lim}}\), one obtains \(\epsilon_{\mathrm{FEL}} < \eta\,\epsilon_{\mathrm{PDU}}\), and therefore the correct condition in Eq.~\eqref{epslims} is \(\epsilon_n < \epsilon_{\mathrm{FEL}}\), with the substitution \(\mu_K \to K_{\mathrm{PDU}}\). Conversely, for \(Q/R > Q/R|_{\mathrm{lim}}\), the relation \(\epsilon_{\mathrm{FEL}} > \eta\,\epsilon_{\mathrm{PDU}}\) holds,  
and the correct condition in Eq.~\eqref{epslims} becomes \(\epsilon_n < \eta\,\epsilon_{\mathrm{PDU}}\), ensuring both \(\mu_K \to K_{\mathrm{PDU}}\) and the theoretical onset of the FEL instability.

\section{Numerical Evaluation}
In this section, a numerical evaluation of the dynamical and radiative properties of the proposed PDU device is presented. The analysis is carried out using the simulation code \textsc{Radyno} \cite{radyno,Frazzitta2023}, which implements fully relativistic particle tracking and radiation post-processing based on the calculation of Liénard--Wiechert potentials \cite{jackson}. Both single-particle and beam radiation spectra are investigated, together with the formation of longitudinal microbunching arising from the interaction between the electron beam and an electromagnetic field during propagation through the PDU. Throughout this section, the simulation of relativistic beam dynamics and radiation emission is treated in a decoupled manner: radiation spectra are computed in post-processing from the particle trajectories, and the evaluation of microbunching is performed by externally injecting an electromagnetic plane wave, which should not be regarded as self-consistently generated by the beam. This approach is adopted with the specific aim of assessing the theoretical capability of the PDU to sustain longitudinal microbunching dynamics analogous to those occurring in conventional undulators. This distinction is especially relevant when comparing the PDU to conventional plasma undulators (CPUs), where electrostatic focusing leads to continuous transverse microbunching \cite{Davoine2018,Hansel2025} rather than the longitudinal density modulation characteristic of undulator-driven free-electron laser processes. 
\begin{figure}[t]
\begin{subfigure}{0.8\linewidth}
\centering
\includegraphics[width=\linewidth]{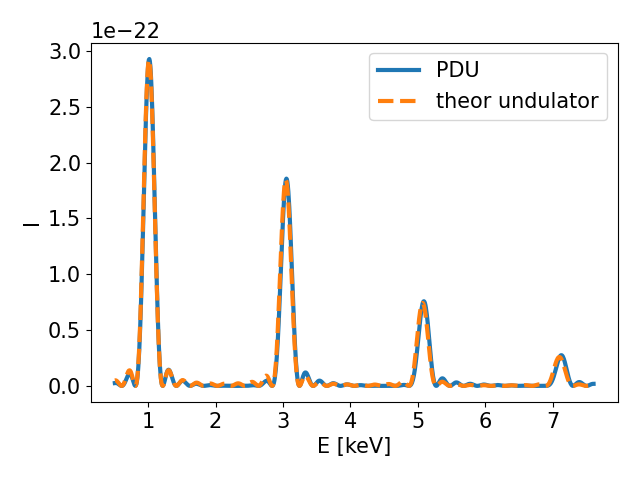}
\caption{}
\label{1part}
\end{subfigure}
\begin{subfigure}{0.8\linewidth}
\centering
\includegraphics[width=\linewidth]{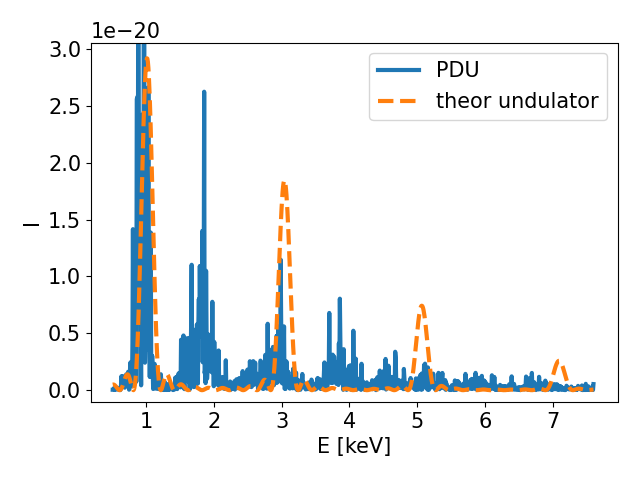}
\caption{}
\label{beam}
\end{subfigure}
\caption{
On-axis radiation spectra for representative PDU parameters.
(a) Single-particle spectrum (solid blue), showing excellent agreement with the theoretical incoherent undulator emission (dashed orange) at the fundamental wavelength and its harmonics.
(b) Spectrum emitted by a matched Gaussian beam (solid blue), exhibiting a slight broadening and deviation from the ideal undulator case (dashed orange) due to the superposition of betatron motion on the forced oscillatory motion, which leads to the appearance of both even and odd harmonics on-axis, associated with mixed-frequency radiation components.
}

\label{onaxis_spec}
\end{figure}

Figure~\ref{onaxis_spec} shows the simulated on-axis radiation spectrum from a single particle (Fig.~\ref{1part}) and from a matched Gaussian beam (Fig.~\ref{beam}) for the parameters \(r_c = \Delta x = 0.5~\mathrm{mm}\), \(h = 3~\mathrm{mm}\), \(I = 10~\mathrm{kA}\), \(K_{\mathrm{PDU}} = 1.12\), \(\gamma = 2000\), \(Q = 100~\mathrm{pC}\), and normalized emittance \(\epsilon_n = 1~\mathrm{mm\,mrad}\). Both are compared to the theoretical incoherent undulator spectrum \cite{jackson} for equivalent beam energy, undulator strength and charge. The fundamental wavelength is computed using the standard expression \(\lambda_1 = \lambda_{\mathrm{PDU}}(1 + K_{\mathrm{PDU}}^2/2)/2 \gamma^2\).
The single-particle spectrum is in perfect agreement with theory, whereas the beam spectrum shows a slight deviation due to the mixture of betatron phases superimposed on the forced oscillation, resulting in an effective single particle oscillation amplitude differing from the nominal \(x_{off}\) given by Eq.~\eqref{x0}. Moreover, because of these phase differences and the relatively large average trajectory slopes, both even and odd harmonics are observed on-axis. This latter feature is further justified by the strong superposition of betatron and undulatory motion, which gives rise to radiation at mixed frequencies corresponding to combinations of the fundamental harmonics $\omega_1 = \omega_{1,\beta}$ and $\omega_2 = \omega_{1,PDU}$, of the form \(\omega = n \omega_1 + m \omega_2 + k (\omega_1 - \omega_2) + l (\omega_1 + \omega_2)\), with integer coefficients $n, m, k,$ and $l$ \cite{Dattoli1992}.

\begin{figure}[t]
\begin{subfigure}{0.99\linewidth}
\centering
\includegraphics[width=\linewidth]{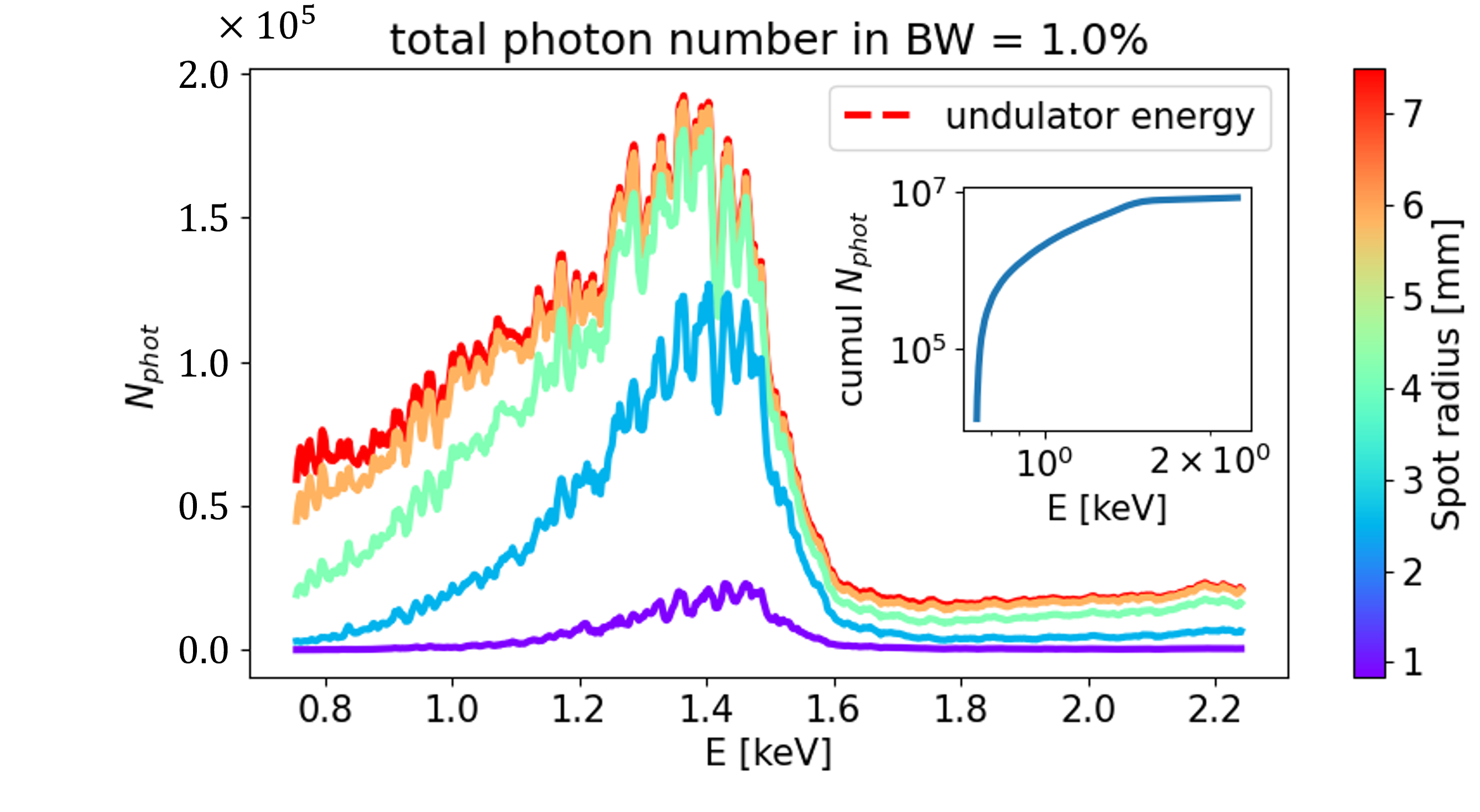}
\caption{}
\label{BW}
\end{subfigure}
\begin{subfigure}{0.9\linewidth}
\centering
\includegraphics[width=\linewidth]{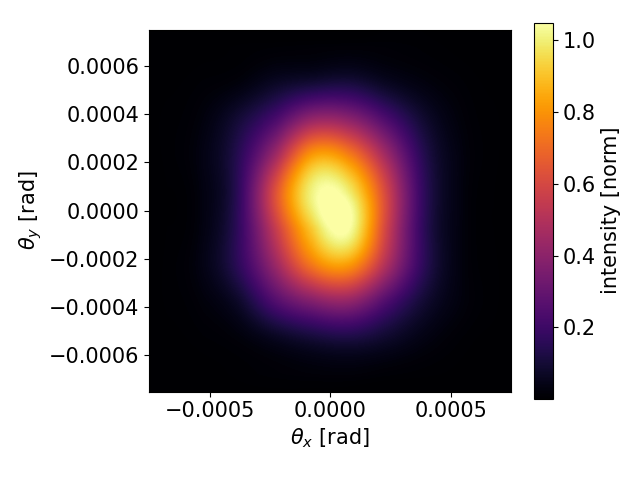}
\caption{}
\label{spot}
\end{subfigure}
\caption{(a) Photon spectrum integrated within a 1\% bandwidth as a function of the radiation spot radius (color scale), for 4~kA discharge current, \(\sigma_r = 4~\upmu\mathrm{m}\), \(K_{\mathrm{PDU}} = 0.45\), and undulator length \(L_{\mathrm{PDU}} = 6~\mathrm{cm}\). The inset plot shows the cumulative photon number. (b) Frequency-integrated radiation spot at 10~m from the source, showing the expected divergence of \(\pm 1/\gamma\) in the \(K_{\mathrm{PDU}} < 1\) regime.}
\label{spectrum}
\end{figure}

Figure~\ref{spectrum} presents an angle-resolved spectrum simulation for parameters similar to those of Fig.~\ref{onaxis_spec}, but with \(I = 4~\mathrm{kA}\), \(\sigma_r = 4~\upmu\mathrm{m}\), \(K_{\mathrm{PDU}} = 0.45\), and device's length \(L_{\mathrm{PDU}} = 6~\mathrm{cm}\) (i.e. ten undulation periods). In Fig.~\ref{BW}, a numerical integration of the photon yield within a 1\% bandwidth is shown as a function of the radiation spot radius at 10~m from the source (color variable), yielding a cumulative incoherent photon number of about \(10^7\) around 1.4~keV. The corresponding frequency-integrated radiation spot, resolved in angle, is shown in Fig.~\ref{spot}, exhibiting the expected divergence consistent with the \(K_{\mathrm{PDU}}<1\) undulator regime (\(\theta_{div}=\pm1/\gamma\) \cite{jackson}).

\begin{figure}[t]
\begin{subfigure}{0.8\linewidth}
\centering
\includegraphics[width=\linewidth]{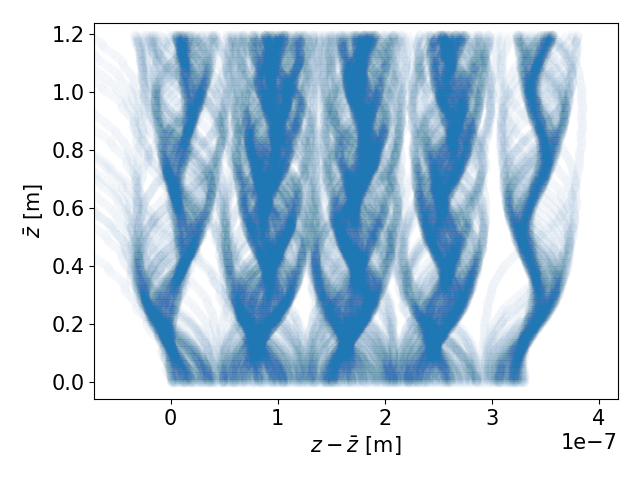}
\caption{}
\label{bunching}
\end{subfigure}
\begin{subfigure}{0.8\linewidth}
\centering
\includegraphics[width=\linewidth]{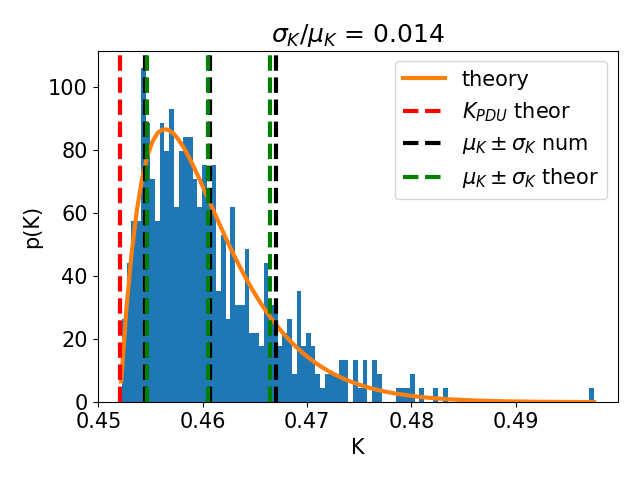}
\caption{}
\label{Kdist}
\end{subfigure}
\caption{
(a) Longitudinal microbunching of a matched electron beam with $\gamma = 200$ and initially uniform longitudinal distribution, driven by an externally injected plane electromagnetic wave of $10^8$ photons at the fundamental PDU wavelength $\lambda_{1,\mathrm{PDU}} = 82\,\mathrm{nm}$. The particle longitudinal coordinate $z$ is shown with respect to the beam centroid $\bar{z}$. Density modulations develop with a periodicity equal to $\lambda_{1,\mathrm{PDU}}$, progressively smoothed by longitudinal betatron nonlinearities.
(b) Distribution of the undulator strength parameter for the same beam (blue binning), compared with the expected Rayleigh $\chi_4$ distribution (solid orange). Numerical estimates of the mean value and standard deviation (dashed black) are compared with the analytical predictions of Eq.~\eqref{Kestim} (dashed green), showing excellent agreement.
}
\label{bunchKdist}
\end{figure}

Figure~\ref{bunching} illustrates two representative numerical results obtained for a matched electron beam with Lorentz factor $\gamma = 200$ and an initially uniform longitudinal distribution. In the first subfigure, an example of longitudinal microbunching is shown. Together with the electron beam propagating in the PDU, a plane electromagnetic wave corresponding to $10^8$ photons at the fundamental emission wavelength $\lambda_{1,\mathrm{PDU}} = 82\,\mathrm{nm}$ is externally injected. The horizontal axis represents the longitudinal position $z$ of the beam particles with respect to the beam centroid $\bar{z}$, while the vertical axis shows the centroid position along the device. Starting from an initially uniform distribution at $\bar{z} = 0$, the particle distribution develops oscillating density modulations, which are progressively smoothed by the nonlinearity of the longitudinal betatron motion. The resulting microbunches are evenly spaced along the longitudinal coordinate, with a separation exactly equal to $\lambda_{1,\mathrm{PDU}}$. Figure~\ref{Kdist} shows, for the same case, the distribution of the undulator strength parameters of the beam particles (blue binning). The numerical distribution is compared with the expected Rayleigh $\chi_4$ distribution (solid orange), showing excellent agreement. In addition, the numerically evaluated mean value and standard deviation of the undulator strength (dashed black lines) are compared with the corresponding analytical estimates derived in Eq.~\eqref{Kestim} (dashed green lines), again demonstrating very good quantitative agreement between simulations and theory.

\section{Conclusions}
A novel plasma-based undulator concept, the \textit{Plasma Discharge Undulator} (PDU), has been presented and theoretically characterized. By introducing a periodic geometric forcing term through a sinusoidal modulation of a plasma discharge, the PDU enables controlled undulatory motion of relativistic electron beams, while freeing the undulation scheme from the need for a high-energy-density external driver such as an intense laser pulse or a particle beam as in conventional plasma undulators (CPUs). The analytical treatment shows that appropriate injection conditions can suppress collective betatron oscillations, yielding a well-defined oscillation at the imposed wavelength $\lambda_{\mathrm{PDU}}$ and significantly reducing the intrinsic $K$-spread typical of CPUs. Design laws and scaling relations have been derived for particle trajectories and radiation emission, and the one-dimensional requirements for free-electron laser (FEL) emission have been evaluated, providing feasibility criteria for FEL operation in the PDU configuration. Moreover, the two limiting expressions for the undulator strength associated with the forced and betatron motion are shown to arise from a more general formulation, highlighting the conditions under which the effective undulator strength converges toward the imposed value $K_{\mathrm{PDU}}$ with a vanishing \(K\)-spread.

Numerical simulations based on fully relativistic particle tracking and radiation post-processing confirm the analytical predictions. Single-particle and beam radiation spectra exhibit narrow-band emission in the millimeter-to-centimeter undulator period regime, extending to the soft X-ray range, with tunable spectral properties and high transverse focusing gradients ($O(\mathrm{kT/m})$) suitable for beam transport and matching. The strong superposition of betatron and undulatory motion leads to the appearance of mixed-frequency radiation components, resulting in the presence of both even and odd harmonics on-axis. Additional numerical studies demonstrate that, under the action of an externally injected electromagnetic field, the PDU is capable of supporting the formation of longitudinal microbunching with periodicity equal to the fundamental emission wavelength $\lambda_{1,\mathrm{PDU}}$, analogous to that observed in conventional magnetic undulators (CUs). 

Concluding with an experimental perspective, the proposed implementation based on geometrically modulated discharge capillaries benefits from the extensive recent development of plasma discharge channels, including nonlinear geometries. These considerations provide strong motivation for the experimental feasibility of current modulation through capillary shaping. Overall, the PDU represents a promising step toward compact, fully plasma-based light sources and accelerator beamline components. Future work will focus on the numerical and experimental validation of the discharge modulation mechanism and on the integration of the PDU with conventional RF and possibly plasma-based acceleration stages.

\section{Acknowledgements}
My sincere gratitude goes to my PhD supervisor, Andrea Renato Rossi, for his patience and for the freedom he has granted me throughout these years; and to the INFN Milan colleagues for the lunch club entertainment. To my co-supervisors: Alessandro Cianchi, for his constant support and for the tennis/life school; and Massimo Ferrario, for the care and determination with which he introduced me to the world of research. My heartfelt thanks also go to Vittoria Petrillo, for her generous and insightful discussions at any time of the day; and to Riccardo Pompili, together with the LNF colleagues, for welcoming me into the experimental campaign for the ABP testing and for never missing a laugh together. Finally, I wish to thank Klaus Floettmann, Hossein Delsim-Hashemi, and all my colleagues at DESY, for showing me how warm Hamburg can be in winter.

My deepest thanks go to my family and friends, for being the best part of my life and for walking along with me every single day.

\bibliography{sample.bib}

\end{document}